\documentclass[twocolumn,superscriptaddress,longbibliography,
	aps,prb,preprintnumbers,10pt]{revtex4-2}

\usepackage[normalem]{ulem}
\usepackage{graphicx}
\usepackage{bm}
\usepackage{color}
\usepackage{epstopdf}
\usepackage{amsmath}
\usepackage{amssymb}
\usepackage{soul}

\usepackage[urlcolor=blue,colorlinks=true,citecolor=blue,linkcolor=blue,pdfstartview={FitH},bookmarks=false]{hyperref}

\usepackage{xcolor}

\graphicspath{{fig/}{./fig/}{.}}

\begin{document}

\title{Chiral phononic and electronic edge modes of EuPtSi}

\author{Issam Mahraj}
\email[e-mail: ]{issam.mahraj@ifj.edu.pl}
\affiliation{Institute of Nuclear Physics, Polish Academy of Sciences, W. E. Radzikowskiego 152, PL-31342 Krak\'{o}w, Poland}

\author{Andrzej Ptok}
\email[e-mail: ]{aptok@mmj.pl}
\affiliation{Institute of Nuclear Physics, Polish Academy of Sciences, W. E. Radzikowskiego 152, PL-31342 Krak\'{o}w, Poland}

\date{\today}

\begin{abstract}
Systems with P2$_{1}$3 symmetry are characterized by the realization of chiral edge modes, propagating in one direction along closed loops around some high symmetry points of the Brillouin zone.
We study the phononic and electronic properties of EuPtSi, which crystallizes with P2$_{1}$3 symmetry.
EuPtSi is also characterized by intriguing magnetic properties, such as the realization of the skyrmion lattice.
Here, using ab initio techniques, we study bulk and slab properties of EuPtSi.
The bulk phononic and electronic band structures exhibit a spin-1 Weyl point and a charge-2 Dirac point at the $\Gamma$ and R points, respectively.
Consequently, the surface states exhibit chiral edge modes.
Such features are present in both the phononic and electronic surface spectra.
The chiral phononic edge mode is associated with the vibration of the atoms in close vicinity, while the chiral electronic surface states correspond to carrier accumulation at the edge of chiral atomic chains.
\end{abstract}

\maketitle

\section{Introduction}

Observation of unusual Weyl fermions in the three-dimensional (3D) noncentrosymmetric transition metal monophosphides~\cite{weng.fang.15,xu.belopolski.15,lv.xu.15,lv.weng.15,yang.liu.15}, has sparked intensive experimental and theoretical studies of exotic states of matter realized in condensed matter systems~\cite{yan.felser.17,armitage.mele.18,manna.sun.18,gao.venderbos.19,hasan.chang.21,lv.qian.21}.
Weyl points, as monopoles of topological charge, give rise to the formation of {\it Fermi arc} connecting two points with opposite charges, which have been observed experimentally in the angle-resolved photoemission spectroscopy (ARPES) experiments~\cite{xu.belopolski.15,lv.xu.15,lv.weng.15,yang.liu.15}. 
Weyl points can also be induced by the Zeeman filed, through the magnetic decoupling of highly-degenerated points in the electronic band structure~\cite{kuroda.tomita.17,morali.batabyal.19,sanchez.chang.20,li.zhang.23}, such as the in the half-Heusler compound GdPtBi~\cite{suzuki.chisnell.16,hirschberger.kushwaha.16,shekhar.kumar.18}.

Exotic states, in the from of the highly degenerated points in the band structure, can also be expected in other compounds~\cite{zhang.gu.20,kumar.yao.20,xian.chapai.20,nie.bernevig.21,jin.liu.21}.
Such states are enforced by the system's symmetry and are realized in electronic~\cite{manes.12,wieder.kim.16,bradlyn.cano.16,yu.zhang.22} as well as in phononic~\cite{jin.wang.18,liu.fu.19,zhang.miao.19,li.liu.21,xie.liu.21,jin.chen.21} systems.
It is worth noting that some crystal symmetries support the realization of the bulk chiral fermions~\cite{manes.12}, with \mbox{spin-1} or \mbox{-3/2} characters.
Such properties were recently reported in the systems with P2$_{1}$3 symmetry, such as monosilicides~\cite{tang.zhou.17,chang.xu.17,pshenayseverin.ivanov.18,bose.narayan.21,takane.wang.19,rao.li.19,sanchez.belopolski.19,cochran.belopolski.23,li.xu.19}, monogermanides~\cite{basak.kobialka.24,barman.mondal.20}, and other related \mbox{compounds~\cite{schroter.pei.19,yao.manna.20,li.xu.19,barman.mondal.20,sessi.fan.20,wang.cui.23,lv.feng.19}.}
These system are also characterized by chiral edge states, propagating along one direction in closeed loops around the points in the Brillouin zone with opposite chirality.
The realization of such exotic states has been reported in phonon dispersion curves and surface states~\cite{miao.zhang.18,zhang.song.18,yang.uphoff.21,jin.hu.22,li.zhang.21}.
In the case of electronic band structure, spin--orbit coupling (SOC) leads to the splitting of the Fermi arc.
Such effect has also been observed experimentally in ARPES \mbox{measurements~\cite{li.xu.19,takane.wang.19,rao.li.19,sanchez.belopolski.19,cochran.belopolski.23,schroter.pei.19,yao.manna.20}.}

Here, we are focused on the EuPtSi crystal with P2$_{1}$3 symmetry.
This compound exhibits interesting magnetic properties.
Initially, EuPtSi was recognized as a trillium lattice with an antiferromagnetic/helimagnetic ground state, having a N\'eel temperature \mbox{$T_\text{N} = 4.1$~K~\cite{franco.prots.17,kaneko.frontzek.19,mishra.ganesan.19}.}
Due to the absence of inversion symmetry, the Dzyaloshinskii--Moriya interaction plays a crucial role in inducing strong magnetic fluctuations~\cite{janoschek.garst.13}.
Indeed, in EuPtSi, strong magnetic fluctuations extend to temperatures well above $T_\text{N}$~\cite{homma.kakihana.19,higa.ito.21}.
Additionally, the magnetic transition at the N\'eel temperature has a pronounced first-order character,  likely due to magnetic frustration~\cite{franco.prots.17,homma.kakihana.19,sakakibara.nakamura.19}.
Moreover, an external magnetic field can induce a magnetic skyrmion lattice (within A- and B-phases)~\cite{kakihana.aoki.18,mishra.ganesan.19,kaneko.frontzek.19,tabata.matsumura.19,tabata.matsumura.19,takeuchi.kakihana.19,kakihana.aoki.19,takeuchu.kakihana.20,hayami.yambe.21,sakakibara.nakamura.21,matsumura.tabata.24}.
EuPtSi is most likely the first rare-earth based intermetallic compound in which the skyrmion lattice was
observed~\cite{kakihana.nishimura.17}.

In this paper, we study the phononic and electronic properties of the chiral EuPtSi crystal.
The system is dynamically stable with P2$_{1}$3 symmetry, as evidenced by the absence of the imaginary soft modes.
The bulk electronic band structure is relatively complex.
The magnetic order of Eu$^{+2}$ is associated with $f$-states located well below the Fermi level.
In practice, these $f$-states do not affect the Fermi surface, which directly shows 3D character of the electrons in the system.
Finally, we study the phononic and electronic surface states of chiral EuPtSi.
Due to the system's symmetry, we found chiral edge modes forming the Fermi arc in both phononic and electronic states. 
These surface states are realized between the highly degenerated bulk states at the $\Gamma$ and R points of the Brillouin zone.
The phononic surface states are associated with atomic vibrations in the close vicinity of the surface.
Similarly, the electronic surface states are associated with the accumulation of carrier distribution at the edge of chiral atomic chains.

\begin{figure*}
\centering
\includegraphics[width=\linewidth]{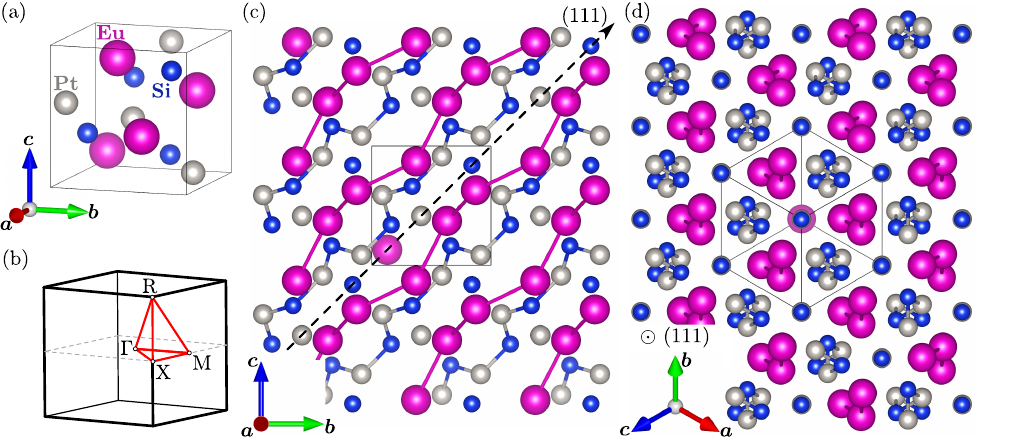}
\caption{
Crystal structure of chiral EuPtSi with P2$_{1}$3 symmetry (a), and the corresponding Brillouin zone with its high symmetry points (b).
Side view of the EuPtSi crystal (c).
Atoms form chiral chains, which are well visible along the (111) direction (d).
\label{fig.crys}
}
\end{figure*}

The paper is organized as follows.
Details of computational techniques are presented in Sec.~\ref{sec.tech}.
In Sec.~\ref{sec.res} we present and discuss of our theoretical results.
In particular, we investigate the bulk properties in Sec.~\ref{sec.bulk} (i.e. the crystal structure and its dynamical stability, the electronic band structure and the Fermi surface, and briefly magnetic properties) as well as the phononic and electronic surface states in Sec.~\ref{sec.ph_surface} and Sec.~\ref{sec.el_surface}, respectively.
Finally, we summarize our paper with the main conclusions in Sec.~\ref{sec.summary}.


\section{Computational details}
\label{sec.tech}

The first-principles density functional theory (DFT) calculations were performed using the projector augmented-wave (PAW) potentials~\cite{blochl.94} implemented in the Vienna Ab initio Simulation Package ({\sc Vasp}) code~\cite{kresse.hafner.94,kresse.furthmuller.96,kresse.joubert.99}.
For the exchange-correlation energy, the generalized gradient approximation (GGA) in the Perdew, Burke, and Ernzerhof (PBE) parametrization was used~\cite{perdew.burke.96}.
The energy cutoff for the plane-wave expansion was set to $600$~eV.
In the case of the Eu $f$-state treated as a valence states, the correlation effects were introduced within the DFT+U scheme, proposed by Dudarev {\it et al.}~\cite{dudarev.botton.98}, with $U = 6$~eV.
The optimization of the lattice constants and atomic positions in the presence of the SOC were performed for the conventional unit cell (containing four formula units), using $8 \times 8 \times 8$ {\bf k}--point grid, following the Monkhorst--Pack scheme~\cite{monkhorst.pack.76}.
As a convergence criterion for the optimization loop, we set the energy change threshold to below $10^{-6}$~eV and $10^{-8}$~eV for the ionic and electronic degrees of freedom, respectively.
The symmetries of the system were analyzed using {\sc FindSym}~\cite{stokes.hatch.05} and {\sc Spglib}~\cite{togo.tanaka.18}, while momentum space analysis was performed with {\sc SeeK-path}~\cite{hinuma.pizzi.17}.

Dynamic properties were calculated using the direct {\it Parlinski--Li--Kawazoe} method~\cite{parlinski.li.97}, implemented in the {\sc Phonopy} package~\cite{togo.chaput.23,togo.23}. 
Within this method, the interatomic force constants (IFCs) are calculated from the Hellmann-Feynman (HF) forces acting on the atoms after displacements of individual atoms inside the supercell.
We performed these calculations using a supercell corresponding to $2 \times 2 \times 2$ unit cells.
During these calculations, a reduced $4 \times 4 \times 4$ {\bf k}-grid was used.

The exact electronic band structure from DFT calculations for the optimized structure was used to find the tight-binding model in the basis of the maximally localized Wannier orbitals~\cite{marzari.mostofi.12,marzari.vanderbilt.97,souza.marzari.01}. 
This was performed using the {\sc Wannier90} software~\cite{mostofi.yates.08,mostofi.yates.14,pizzi.vitale.20}.
In our calculations, we used an $8 \times 8 \times 8$ $\Gamma$-centered ${\bm k}$-point mesh.
Here, calculations were perform for Eu $f$ electrons treated as core states.
As a starting point, we used $d$ orbitals for Eu, $s$ and $d$ for Pt, and $s$ and $p$ orbitals for Si.
Finally, we found a $60$-orbital, $120$-band model reproducing the electronic band structure over a wide range of energies.
Similarlly, the obtained IFCs were used to generate the phonon tight-binding Hamiltonian.
Both electronic and phononic tight-binding models were used to study the surface states.
The surface Green's function for a semi-infinite system~\cite{sancho.sancho.85} was calculated using {\sc WannierTools}~\cite{wu.zhang.18}.

\begin{figure}[!t]
\centering
\includegraphics[width=\linewidth]{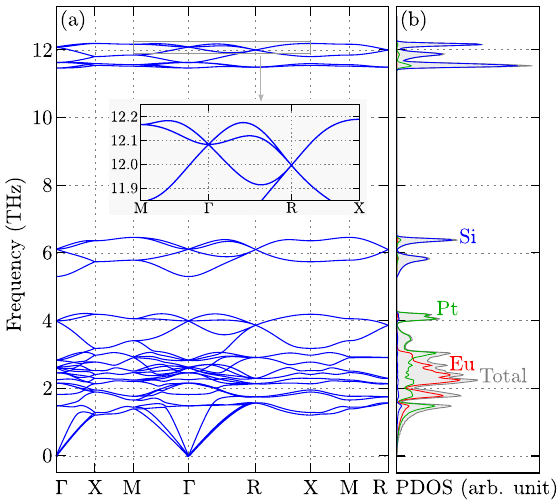}
\caption{
The phonon dispersion curves (a) and density of states (b).
The inset presents a zoomed view of the phonon dispersion curves around the $\Gamma$ and R points in the high-frequency range.
\label{fig.ph_band}
}
\end{figure}

\begin{figure*}
\centering
\includegraphics[width=\linewidth]{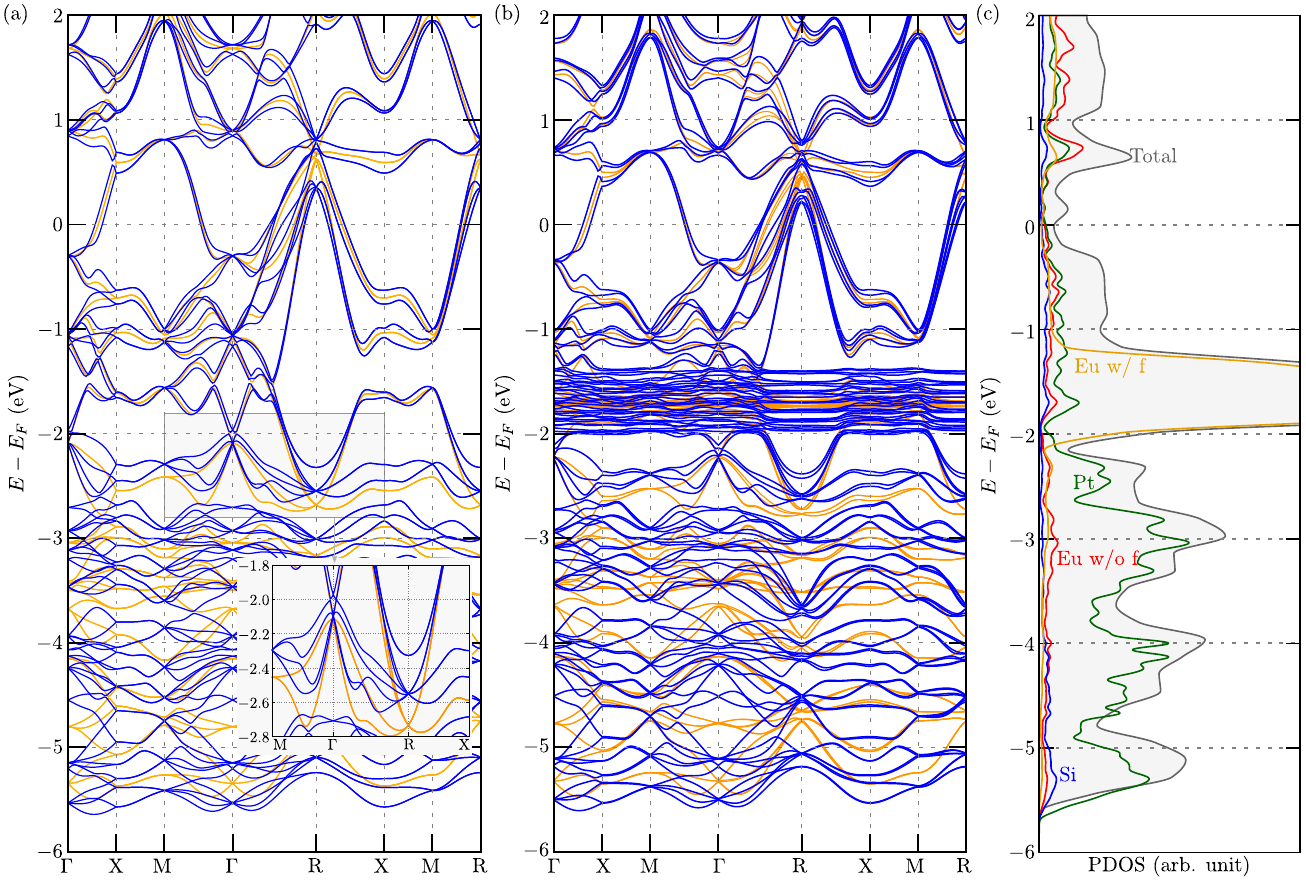}
\caption{
The electronic band structure of EuPtSi with Eu $f$-states treated as core states (a) and valence states (b).
The inset in panel (a) presents a zoomed view of the bands around the spin-1 Weyl and charge-2 Dirac points at $\Gamma$ and R points, respectively.
Results are shown in the absence and presence of spin--orbit coupling (orange and blue lines, respectively) for $U = 6$~eV.
The electronic partial density of states is shown in panel (c).
\label{fig.el_band}
}
\end{figure*}

\begin{figure}[!b]
\centering
\includegraphics[width=0.9\linewidth]{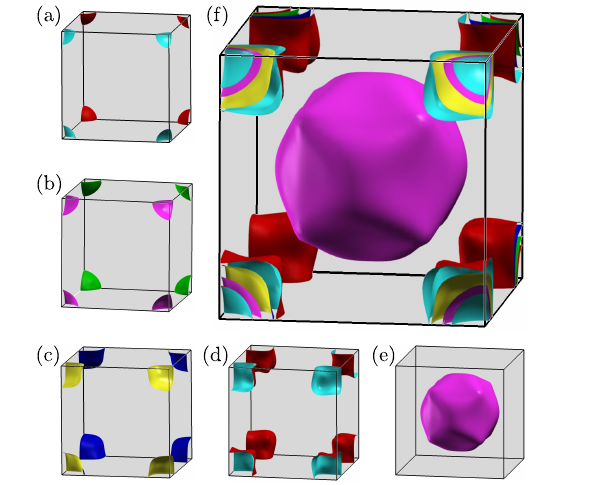}
\caption{
The Fermi surface of EuPtSi. 
Panels from (a) to (e) show separate Fermi pockets, while (f) show the ``full'' Fermi surface.
\label{fig.fs}
}
\end{figure}

\begin{figure*}
\centering
\includegraphics[width=\linewidth]{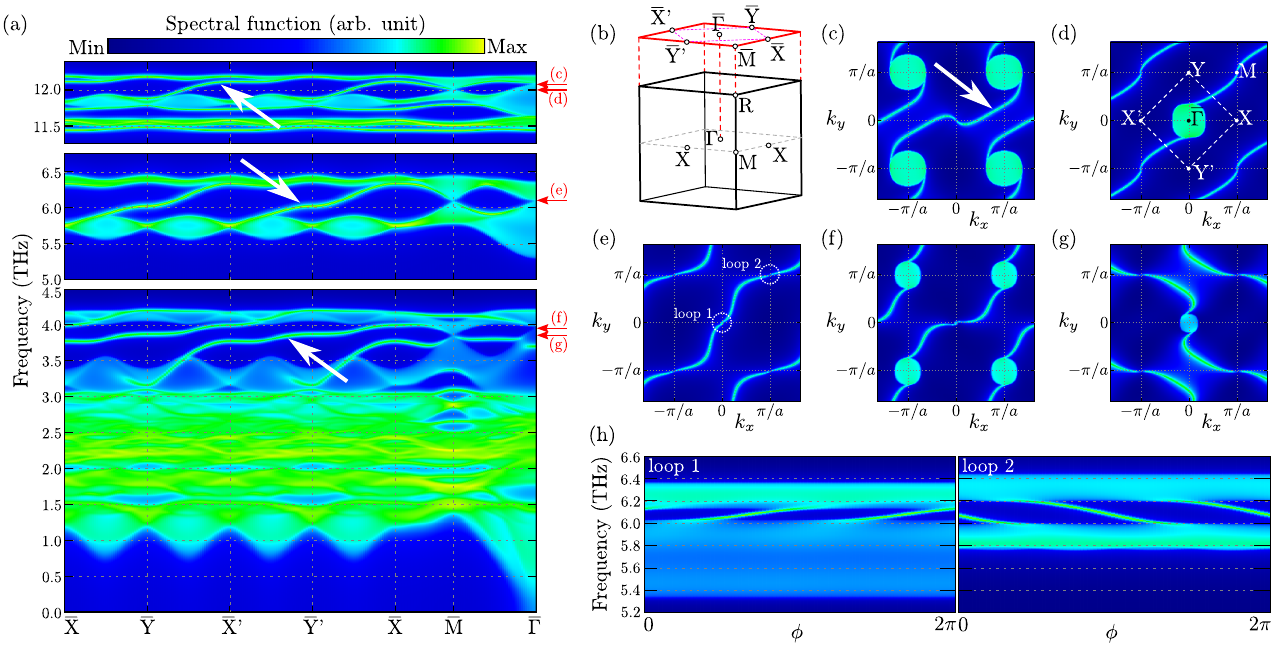}
\caption{
The phononic spectral function for the (001) surface along the path presented in (b).
Projection of the bulk Brillouin zone  onto the surface Brillouin zone is shown in (b).
The constant frequency contours for the cuts marked by red arrows in (a) are shown for
$12.08$~THz (c), $12.00$~THz (d), $6.10$~THz (e), $3.96$~THz (f), and $3.86$~THz (g).
The phononic surface spectral function along the loops presented in (e) is shown.
The loop 1 and 2 correspond to loops around the $\bar{\Gamma}$ and $\bar{\text{M}}$ points, respectively.
\label{fig.ph_ss}
}
\end{figure*}


\section{Results and discussion}
\label{sec.res}

\subsection{Bulk properties}
\label{sec.bulk}

{\it Crystal structure.}---
EuPtSi crystallizes in the cubic LaIrSi-type structure, with P2$_{1}$3 symmetry (space group No.~198).
The crystal structure is presented in Fig.~\ref{fig.crys}.
The system is characterized by three twofold screw rotations along Cartesian directions. 
Additionally, there is one threefold rotation symmetry axis along (111) direction.
Firstly, such symmetries affect the phononic and electronic bulk band structures, which will be discussed in the following paragraphs.
Secondly, the realization of chiral atomic chains is expected. 
Indeed, such chains are well visible along the (111) direction [see Fig.~\ref{fig.crys}(d)].
Contrary to binary compounds, like monosilicides~\cite{tang.zhou.17,chang.xu.17,pshenayseverin.ivanov.18,bose.narayan.21,li.xu.19,takane.wang.19,rao.li.19,sanchez.belopolski.19,cochran.belopolski.23} or monogermanides~\cite{barman.mondal.20,basak.kobialka.24}, EuPtSi realizes two types of chains: (i) chains of Eu atoms, and (ii) chains constracuted by Pt and Si atoms.
Additionally, some atoms (of all types) are located outside the chains [see, e.g., the atoms in the center of the hexagonal cell in Fig.~\ref{fig.crys}(d)].

All atoms in EuPtSi are located in the $4a$ Wyckoff position ($x$,$x$,$x$), thus the unit cell contains four chemical formulas.
After optimization the crystal structure, we found $a = 6.429$~\AA, which is in excellent agreement with the experimental value of $a = 6.436$~\AA~\cite{franco.prots.17}.
Additionally, the ``free'' parameters $x$ describing atomic positions were determined to be $0.3665$, $0.0829$, and $0.6634$ for Eu, Pt, and Si, respectively, while the experimentally reported values are $0.3700$, $0.0863$, and $0.6611$~\cite{franco.prots.17}.

{\it Phonon spectra.}---
The phonon dispersion curves are presented in Fig.~\ref{fig.ph_band}.
Due to the absence of imaginary soft modes, EuPtSi is stable in the P2$_{1}$3 structure in dynamical sense.
In the phonon spectra, we can recognize a few groups of bands: 
(i) a low-frequency range (below $5$~THz), containing, e.g., acoustic branches;
(ii) a middle-frequency range (around $6$~THz);
and (iii) a high frequency range (around $12$~THz).
The acoustic branches around the $\Gamma$ point possess nearly ideal linear dispersion and are well hybridized with optical branches above $2$~THz.
Optical branches in the middle and high-frequency ranges are relatively narrow.

The formation of such groups of bands is also reflected in the phonon density of states (phDOS), presented in Fig.~\ref{fig.ph_band}(b).
Due to their atomic masses, the vibrations associated with the heavies atoms (Eu and Pt, with comparable atomic masses) occur mostly in low-frequency range.
In practice, the small contribution of Si vibrations to the phDOS in the low-frequency range is realized only within the acoustic branches.
Moreover, the middle and high-frequency phonon modes are associated with vibrations of Si atoms.
Above $5$~THz, there is only small contribution of Pt atoms to the phDOS (at the top of the middle-frequency group and the bottom of the high-frequency group).
This is associated with the vibration of Pt and Si atoms, which are well coupled within the same Pt-Si chains.
A similar situation was observed in BaPtGe, also with P2$_{1}$3 symmetry, where high-frequency branches (with Pt and Ge contributions) are separated from another modes (mixed by vibrations of all atoms)~\cite{li.zhang.21}.

The crystal symmetries gives rise to threefold representations at the $\Gamma$ point and fourfold representations at the R point [see inset in Fig.~\ref{fig.ph_band}(a)].
As consequence, in the phononic band structure, we can recognize a spin-1 Weyl point (at $\Gamma$) and an effective charge-2 Dirac point (at R), which are special cases of double Weyl points~\cite{zhang.song.18}.
Both topological points possess opposite non-zero Chern number ($\pm 2$), and their total Chern numbers cancel exactly within the  Brillouin zone.
The spin-1 Weyl point is formed by the crossing of three branches, where two of them form Dirac cones, while the third forms a nearly-flat band.
The crossing of all these bands at the same frequency gives rise to the triple degeneracy at the $\Gamma$ point. 
In contrast, the charge-2 Dirac point is formed directly by the fourfold degenerate states at R.
In this case, a doubly-degenerate Dirac cone is realized around the R point.
Both points, $\Gamma$ and R, exist due to the crystal symmetry and time-reversal symmetry.
In practice, the spin-1 Weyl and charge-2 Dirac points can be found in the phonon spectra over a wide range of frequencies.

{\it Electronic and and Fermi surface.}---
The electronic band structure for Eu $f$ electrons treated as core and valence states are presented in Fig.~\ref{fig.el_band}(a) and~\ref{fig.el_band}(b), respectively.
As we can see, the localized Eu $f$-states are located far below the Fermi level (around $-1.75$~eV).
When $f$-states are treated as valence states, we observe additional band splitting due to the exchange field acting on the other states.
However, the $f$-states do not strongly affect the electronic band structure [cf.~blue lines in Fig.~\ref{fig.el_band}(a) and~\ref{fig.el_band}(b)].
Moreover, the electronic band structure of EuPtSi with $f$ electrons treated as a core states is similar to that reported earlier for SrPtSi (without $f$-states)~\cite{kakihana.aoki.19}.

In the absence of SOC, the electronic band structure features a spin-1 Weyl point and an effective charge-2 Dirac point at the $\Gamma$ and R points, respectively [e.g., orange lines in the inset of Fig.~\ref{fig.el_band}(a)].
For instance, the spin-1 Weyl point is visible around $-2.1$~eV, while the charge-2 Dirac point is around $-2.75$~eV.
Introducing SOC leads to the decoupling of spin-1 state from a spin-degenerate orbital triplet into a doubly-degenerate point and fourfold-degenerate states (with a Chern number of $+4$).
Similarly, at the R point, SOC leads to the decoupling into a sixfold fermionic point (with a Chern number of $-4$) and a trivial doubly-degenerate point.
Such effect is shown in the inset of Fig.~\ref{fig.el_band}(a).

We should also note that relatively strong spin--orbit effects [cf.~orange and blue lines in Fig.~\ref{fig.el_band}(a) and~\ref{fig.el_band}(b)] are observed well below the Fermi level (below $-2$~eV).
The density of states [Fig.~\ref{fig.el_band}(c)] uncover reveals a strong Pt $d$-states contribution in this energy range.
Heavy atoms introduce strong SOC, and indeed, Pt has the largest spin--obit contribution within $d$-states ($60$~meV) and $p$-states ($38$~meV).
In contrast, for Eu, SOC contributions are $140$~meV, $0.3$~meV, $13$~meV for $p$-, $d$-, and $f$-states, respectively, while spin--orbit affect of Si is negligible ($< 0.2$~meV).

The Fermi surface of EuPtSi (Fig.~\ref{fig.fs}) exhibits a distinctly three-dimensional character of electrons.
The Fermi surfaces take the shape of deformed spheres centered at $R$ [Fig.~\ref{fig.fs}(a)--(d)] or $\Gamma$ [Fig.~\ref{fig.fs}(e)], and are much simpler than those in pure Pt~\cite{xu.digennaro.20}, while similar to those reported for PdGa~\cite{maulana.li.21} or PtGa~\cite{schwarze.uhlarz.22}.
Such shape of the Fermi pockets was confirmed experimentally, through de Haas--van Alphen oscillations measurements~\cite{kakihana.aoki.19}.
The Fermi pocket centered on $\Gamma$ has electron-like character, while the one centered on $R$ has hole-like character.

The density of states is mainly due to contributions from Eu $d$--$f$, Pt $d$- (mostly below $-2$~eV), and Si $p$-states.
It is noteworthy that there are no Eu $d$ electrons in initial calculations (Eu has a valence configuration of $5p^{6} 6s^{2}$ when $f$-states are treated as core states), but Eu $d$ electrons contribute to the density of states at the Fermi level in the final electronic state.

{\it Magnetic properties.}---
The magnetic properties of EuPtSi mentioned in the Introduction are associated with the Eu $f$-states.
As we mentioned in the previous paragraph, these states are located well below the Fermi level. 
However, the strong magnetic moment of Eu$^{+2}$ due to the exchange field also affects other states.
In the self-consistent calculation, we found small residual magnetic moments ($< 2$~m$\mu_\text{B}$) on Pt and Si as well.
This is visible in the band structure as additional small band splittings [visible in Fig.~\ref{fig.el_band}(b)], even far from the Eu $f$-states.

For simplicity, in our calculations containing the Eu $f$-states, we assumed an antiferromagnetic (AFM) order.
Nevertheless, the experimentally observed helimagnetic order should not change the band structure significantly.  
The experimentally observed saturation magnetic moment is close to the nominal value of $7$~$\mu_\text{B}$~\cite{franco.prots.17}, as expected for Eu$^{+2}$ with spin $7/2$~\cite{singh.dan.23,swain.sarkar.24}.
From the self-consistent calculations, we determined the Eu magnetic moment to be $6.96$~$\mu_\text{B}$.
The occupied Eu $f$-states are located far below the Fermi level [from $-2.0$~eV to $-1.5$~eV, cf. Fig.~\ref{fig.el_band}(a) and~\ref{fig.el_band}(b)].
As consequence, the main features of the electronic band structure are the same independently of the Eu $f$ electrons.

The magnetic order can sometimes strongly affect on the phonon spectra.
For example, phonon spectra renormalization was reported in the chiral FeSi crystal~\cite{krannich.sidis.15}.
However, contrary to the EuPtSi system, in FeSi, the magnetic order is associated with the partially occupied Fe $d$-states located around the Fermi level.
Thus, the magnetic moments of Fe strongly change the electronic band structure and affect the phonon spectra indirectly due to the modification of the IFCs.
In our case, the electronic band structure around the Fermi level is practically the same in the absence and presence of Eu $f$-states.
Therefore,  investigating the phonon spectra with Eu $f$-states treated as core states is reasonable.

\begin{figure*}
\centering
\includegraphics[width=\linewidth]{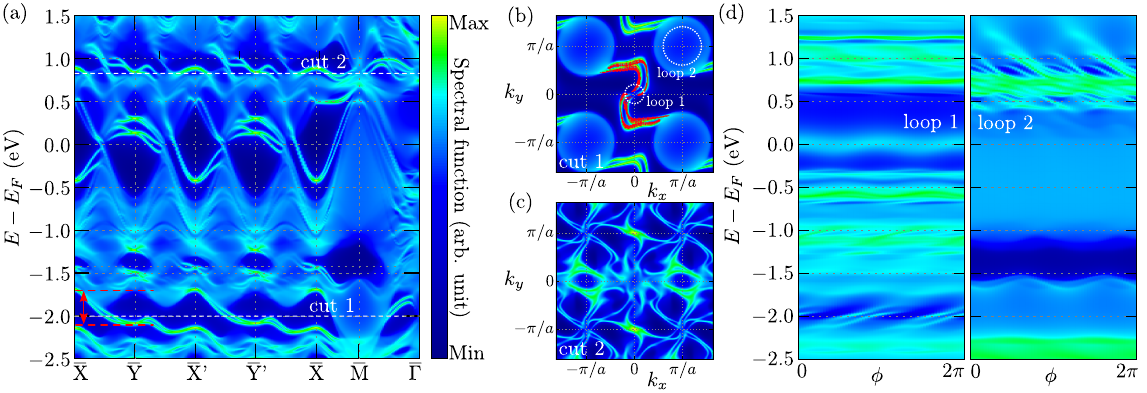}
\caption{
The electronic spectral function for the (001) surface along the path presented in Fig.~\ref{fig.ph_ss}(b).
The constant energy contour for two cuts shown in (a): cut 1 at $-2$~eV (b), and cut 2 at $0.84$~eV (c).
The spectral function along loop 1 and loop 2 (d), around the $\bar{\Gamma}$ and $\bar{\text{M}}$ points, respectively.
\label{fig.el_ss}
}
\end{figure*}

\subsection{Phononic surface states}
\label{sec.ph_surface}

As we mentioned in the previous section, due to the absence of SOC-like effects, the phonon bulk dispersion curve realizes ideal spin-1 Weyl points and charge-2 Dirac points at $\Gamma$ and R, respectively.
Consequently, the emergence of chiral surface states can be expected.
In this section, we will focus on the phononic surface states of EuPtSi.

Our results regarding the phonon surface states for the (001) surface are presented in Fig.~\ref{fig.ph_ss}.
The chiral phononic edge modes should connect the $\Gamma$ and R points of the bulk 3D Brillouin zone projected onto the surface two-dimensional (2D) Brillouin zone, as shown in Fig.~\ref{fig.el_band}(b).
The bulk $\Gamma$ and R points are projected onto the surface $\bar{\Gamma}$ and $\bar{\text{M}}$ points, respectively.
All surface states should be visible as states crossing any path between $\bar{\Gamma}$ and $\bar{\text{M}}$ points, e.g., the $\bar{\text{X}}$--$\bar{\text{Y}}$ path [Fig.~\ref{fig.ph_ss}(b)]. 
Indeed, the spectral function clearly shows surface states crossing paths close to $\bar{\Gamma}$ and $\bar{\text{M}}$ points, such as the $\bar{\text{X}}$--$\bar{\text{Y}}$--$\bar{\text{X'}}$--$\bar{\text{Y'}}$--$\bar{\text{X}}$ path [Fig.~\ref{fig.ph_ss}(a)].  
In practice, the chiral edge modes can be found in the whole range of frequencies, in the form of separate states connecting the two (bottom and top) ``bands'', for example, the states marked by white arrows in Fig.~\ref{fig.ph_ss}(a).

We also present the path $\bar{\text{X}}$--$\bar{\text{M}}$--$\bar{\Gamma}$.
As we can see, at $\bar{\Gamma}$ and $\bar{\text{M}}$ we find Dirac-cone-like structures at the frequencies marked by red arrows in Fig.~\ref{fig.ph_ss}(a). 
Indeed, these points contain states from the highly degenerate bulk points $\Gamma$ and R.
In the low- and high-frequency ranges, the Dirac-cone-like crossings are realized in close pairs of frequencies (i.e., $12.08$~THz and $12.00$~THz, $3.96$~THz and $3.86$~THz).
Contrary to this, in the intermediate frequencies, both points coincide at the same frequency of $6.10$~THz.

The structure of the spectral function allows for the simple identification of the Fermi arcs formed by the chiral edge modes between $\bar{\Gamma}$ and $\bar{\text{M}}$ points [see Fig.~\ref{fig.ph_ss}(c)--(g)].
Figures~\ref{fig.ph_ss}(c),~\ref{fig.ph_ss}(e), and~\ref{fig.ph_ss}(f) correspond to the frequencies of the Dirac-cone-like crossings at $\bar{\Gamma}$, while figures~\ref{fig.ph_ss}(d),~\ref{fig.ph_ss}(e), and~\ref{fig.ph_ss}(g) correspond to the frequencies of the Dirac-cone-like crossings at $\bar{\text{M}}$.
In such constant frequency cross-sections, the Dirac-cone-like crossing is visible in the form of single point at $\bar{\Gamma}$ or $\bar{\text{M}}$.
When the frequencies of crossings at both points are not the same, the bulk states are visible at one of them in the form of a nearly ideal circle.
For example, in Fig.~\ref{fig.ph_ss}(c) at $\bar{\Gamma}$ we see the Dirac-cone-like crossing, while at $\bar{\text{M}}$ we see the bulk states.
Here, the Fermi arc (marked by white arrow) connects both states.
In the case of Fig.~\ref{fig.ph_ss}(e), the Fermi arc is realized between two Dirac-cone-like crossings.
The realization of the Fermi arc between $\bar{\Gamma}$ and $\bar{\text{M}}$ by the chiral edge modes is clearly visible in all cases presented in panels Fig.~\ref{fig.ph_ss}(c)--(g).

Finally, the direct presentation of the chiral edge modes can be visualized by the spectral function around the closeed loops around $\bar{\Gamma}$ and $\bar{\text{M}}$ points [as shown by dashed white circles in Fig.~\ref{fig.ph_ss}(e)].
Results are presented in Fig.~\ref{fig.ph_ss}(h), where the left and right panels correspond to loop 1 and 2, i.e., around $\bar{\Gamma}$ and $\bar{\text{M}}$ points, respectively.
We present here the intermediate frequency range. 
As we mentioned earlier, the chiral modes are visible in the form of edge modes propagating along one direction.
Moreover, the surface states connect the lower and upper parts of the bands (below and above $6.1$~THz), and exist inside the gap between these parts.

\subsection{Electronic surface states}
\label{sec.el_surface}

Now, we will discuss the corresponding surface states problem in the context of the electronic spectrum.
Contrary to the phonon spectrum, the electronic band structure exhibits additional band splitting due to the strong SOC. 
As we mention earlier, the SOC leads to decoupling of the spin-1 Weyl and charge-2 Dirac points. 
The Weyl point decouples into a doubly degenerate point and fourfold degenerats states (with Chern number $+4$), while the Dirac point decouples into a sixfold fermionic point (with Chern number $-4$) and a trivial doubly degenerate point.

Compared to the phonon dispersion curves, the electronic band structure has a much more complex form.
Moreover, the SOC-induced decoupling also affects the chiral edge modes.
Results of the spectral function obtained for the electronic spectrum of EuPtSi for (001) are presented in Fig.~\ref{fig.el_ss}.
Along the $\bar{\text{X}}$--$\bar{\text{Y}}$--$\bar{\text{X'}}$--$\bar{\text{Y'}}$--$\bar{\text{X}}$ path, several surface states are visible [see Fig.~\ref{fig.el_ss}(a)].
However, not all surface states exhibit chiral features.
For example, a chiral edge mode is well visible much below the Fermi level (around $-2$~eV).
Moreover, for other energies, the chiral edge mode can be mixed with other edge modes.
To illustrate both cases, we show constant energy cross-sections at energies of $-2$~eV and $0.83$~eV [cuts 1 and 2 in Fig.~\ref{fig.el_ss}(a)].
Results for cut 1 and 2 are presented in Fig.~\ref{fig.el_ss}(b) and~\ref{fig.el_ss}(c), respectively.
For energy $-2$~eV (cut 1), clearly decoupled chiral edge modes connect the $\bar{\Gamma}$ and $\bar{\text{M}}$ points.
Along the Fermi arc, the SOC-induced unconventional spin texture is evident, as indicated by the red arrows in Fig.~\ref{fig.el_ss}(b).
For energy $0.83$~eV (cut 2), the chiral edge mode is less prominent.
Surface states are also observed around $\bar{\text{X}}$ and $\bar{\text{Y}}$ [Fig.~\ref{fig.el_ss}(c)].

Similarly, as for the phonons, we also calculated the spectral function along loops around the $\bar{\Gamma}$ and $\bar{\text{M}}$ points [Fig.~\ref{fig.el_ss}(d)].
In this case, the chiral edges are visible within some energy ranges.
However, comparing both loops reveals a much more complicated spectrum than in the phonon case.
When the chiral edge mode is visible at some energies in one loop, the bulk states are visible in the second loop [cf.~left and right panels in Fig.~\ref{fig.el_ss}(d)].
It should also be noted that the loop 2 is twice the size of loop 1.
Regardless of this, the chiral edge modes within loop 2 are much more mixed with other states.

\begin{figure}[!t]
\centering
\includegraphics[width=\linewidth]{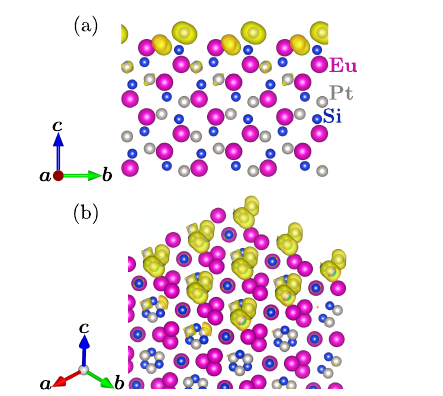}
\caption{
Side view (a) and view along the (111) direction (b) of the EuPtSi crystal, together with the calculated charge density distribution for states within energy window from $-2.1$~eV to $-1.7$~eV [marked by red dashed lines in panel Fig.~\ref{fig.el_ss}(a)].
\label{fig.el_chg}
}
\end{figure}

\section{Summary}
\label{sec.summary}

In this paper, we investigate EuPtSi with P2$_{1}$3 chiral symmetry.
We present complex study of the phononic and electronic properties of the bulk as well as the slab structure with a (001) surface.
We show that EuPtSi with a P2$_{1}$3 structure is dynamically stable due to the absence of imaginary soft modes.
The magnetic order of Eu$^{+2}$ does not strongly affect the electronic band structure.

In the phononic bulk dispersion curves, due to the absence of SOC-like effects, ideal spin-1 Weyl and charge-2 Dirac points are realized at the $\Gamma$ and R points, respectively. 
This is possible as a consequence of the coexistence of three twofold screw rotations along Cartesian directions and one threefold rotational symmetry axis around the (111) direction.
Moreover, the non-zero Chern numbers $\pm 2$ at the corresponding points of the 2D surface Brillouin zone (i.e., $\bar{\Gamma}$ and $\bar{\text{R}}$ points) allow for the realization of the chiral edge modes connecting these points.
For the constant frequency cross-section, such a chiral edge mode is visible in the form of a Fermi arc, connecting $\bar{\Gamma}$ and $\bar{\text{R}}$ points.
Similarly, the chiral edge mode is visible in the phonon spectrum as a mode propagating along one direction in closed loop around the $\bar{\Gamma}$ or $\bar{\text{R}}$ point.

The situation is more complicated in the case of the electronic spectrum. 
Due to the SOC effect, the ideal spin-1 Weyl and charge-2 Dirac points are split and no longer observed.
However, there are still points with non-zero Chern numbers $\pm 4$ at $\bar{\Gamma}$ and $\bar{\text{R}}$ points.
From this, the edge states are realized in the form of split chiral edge modes. 
The main features of such modes are the same as in the case of the phononic chiral edge modes.
Additionally, an unconventional spin texture exists on the surface states.

The phononic and electronic chiral edge modes directly visualize the ``value'' of the Chern number at $\bar{\Gamma}$ and $\bar{\text{R}}$ points.
In both cases, the number of the Fermi arcs connecting $\bar{\Gamma}$ and $\bar{\text{R}}$ points is equal to the absolute value of Chern number, while the chirality corresponds to the sign.
Thus, in the phononic spectrum, there are only two Fermi arcs at $\bar{\Gamma}$ and $\bar{\text{R}}$ points, while due to the SOC, there are four Fermi arcs in the case of electronic spectrum.

The surface states should be associated with edge features. 
For instance, the phononic edge modes are associated with vibrations of the atoms in close vicinity of the surface.
Such behavior was reported, for example, in 2D graphen~\cite{li.wang.20}.
Here, the chiral vibrations of the atoms on the edge can be realized in different chains depending on the frequencies range.
The vibrations within Pt-Si chains are well visible in the phonon spectrum for the frequencies with the largest Si atoms contributions.
Thus, for the phononic chiral edge modes, the vibrations on the surface are realized mostly by Si atoms within Pt-Si chains.
In the case of Eu chains, the edge modes are hard to extract from the bulk spectrum due to the location of Eu modes in the low-frequency range.

Similarly, in the case of the electronic surface state, some states localization can be expected. 
Indeed, this is well visible in the calculated charge density distribution for states within the energy window from $-2.1$~eV to $-1.7$~eV [marked by red dashed lines in panel Fig.~\ref{fig.el_ss}(a)], i.e., for the range of energies corresponding to the electronic chiral edge modes.
From this, we can see directly that carrier accumulation is realized within Pt-Si chains, only on atoms around the surface. 
The absence of  localization on Eu atoms can be consequence of the relatively large distance between atoms inside the Eu chain.
Distances between atoms in Pt-Si chains are around $2.33$~\AA, while in Eu chains, they are around $\sim 3.94$~\AA.
The chiral surface states around the Fermi level do not exhibit a tendency for strong localization around the surface due to the strong mixing with the bulk states.
However, the corresponding states still remain along Pt-Si chains.

The problem of the experimental observation of the chiral edge modes still remains open.
In the case of the phononic surface states,  inelastic x-ray scattering (IXS) or inelastic neutron scattering (INS) can be used.
These techniques were used in the cases of FeSi~\cite{miao.zhang.18} and BaPtGe~\cite{li.zhang.21} to study the charge-2 Dirac point at the R point.
Similarly, INS was used to study the Chern numbers of the phonon branches in MnSi and CoSi~\cite{jin.hu.22}. 
We can assume that IXS and INS can also be helpfully in the study of chiral edge modes when the edge modes are well separated from the bulk states.
Similarly, in the case of the electronic surface states, the ARPES technique can be used.
In this case, the observation of the Fermi arc can confirm the realization of the SOC--split electronic chiral edge modes, as in the cases of CoSi~\cite{rao.li.19,li.xu.19,sanchez.belopolski.19}, RhSi~\cite{sanchez.belopolski.19,cochran.belopolski.23}, RhSn~\cite{li.xu.19}, and PtAl~\cite{schroter.pei.19}.

In summary, the interesting properties of EuPtSi are not limited to its magnetic properties.
The chiral P2$_{1}$3 symmetry allows the realization of highly degenerate points and exotics states in the phononic and electronic spectra.
The realization of helimagnetic order and (phononic and electronic) chiral edge modes makes EuPtSi an excellent platform to study the interplay between different topological features.


\begin{acknowledgments}
Some figures in this work were rendered using {\sc Vesta}~\cite{momma.izumi.11} and {\sc XCrySDen}~\cite{kokalj.99} software.
We kindly acknowledge support by National Science Centre (NCN, Poland) under Project No.~2021/43/B/ST3/02166 (A.P).
\end{acknowledgments}

\bibliography{biblio.bib}


\end{document}